# A PM2.5 concentration prediction framework with vehicle tracking system: From cause to effect

Chuong D.Le
*Department of Computer Science and Engineering*
*Vietnamese German University*
Binh Duong, Vietnam
15983@student.vgu.edu.vn

Hoang V. Pham
*Department of Computer Science and Engineering*
*Vietnamese German University*
Binh Duong, Vietnam
14708@student.vgu.edu.vn

Duy A. Pham
*Computer Science Department*
*Hochschule Bonn-Rhein-Sieg*
Sankt Augustin, Germany
duyanhpham@outlook.com

An D. Le
*Department of Electrical and Computer Engineering*
*University of California San Diego*
La Jolla, USA
d0le@ucsd.edu

Hien B. Vo
*Department of Electrical and Computer Engineering*
*Vietnamese German University*
Binh Duong, Vietnam
hien.vb@vgu.edu.vn

*Abstract*— Air pollution is an emerging problem that needs to be solved especially in developed and developing countries. In Vietnam, air pollution is also a concerning issue in big cities such as Hanoi and Ho Chi Minh cities where air pollution comes mostly from vehicles such as cars and motorbikes. In order to tackle the problem, the paper focuses on developing a solution that can estimate the emitted PM2.5 pollutants by counting the number of vehicles in the traffic. We first investigated among the recent object detection models and developed our own traffic surveillance system. The observed traffic density showed a similar trend to the measured PM2.5 with a certain lagging in time, suggesting a relation between traffic density and PM2.5. We further express this relationship with a mathematical model which can estimate the PM2.5 value based on the observed traffic density. The estimated result showed a great correlation with the measured PM2.5 plots in the urban area context.

## I. INTRODUCTION

Traffic congestion and transportation-related environmental pollution are identified as the severe problems in many countries all over the world, especially in the developing countries, and Vietnam is also not an exception. Ho Chi Minh is one of the largest cities in Viet Nam and has been subsequently expanded to adapt to the growing population. The population of Ho Chi Minh City in 1989 was 2.5 million, increasing at a rate of 4 percent until 1999 and reaching a population of 3.9 million in 1999. The average annual change rate increased to 6.3 per cent, and Ho Chi Minh City's population reached 10.1 million in 2015 [5]. Moreover, the problem relating to internal immigration brings a huge challenge for Ho Chi Minh city. From 2010-2015 36.2% of migration was rural-urban, 31.6% urban-urban, 19.6% rural-rural, and 12.6% urban-rural. Rural-urban migration has fueled rapid urbanization, with the urban population growing by 3.4% per year in comparison to 0.4% in rural Viet Nam [6]. Consequently, over a decade, the number of motorcycles in HCMC has reached an alarming rate with approximately 7 million motorcycles in 2014 [7]. The result leads to the increase of PM2.5 rate as well and causes many negative effects on human health.

PM2.5 (particles less than 2.5 micrometers in diameter) is an air pollutant which has a small diameter. However, it has large surface areas which may be capable of carrying various toxic stuff, passing through the filtration of nose hair, penetrate deeply into our lungs and consequently impair the lung function. Evidently, research published in [8] indicates that smog emanating from motor vehicles is typically the result of excessive concentrations of tiny particles (particle size less than or equal to 2.5, referred to as PM2.5).

Therefore, in this study, we developed a vehicle counting algorithm and forecasting system to analyze the amount of PM2.5 compared to the number of vehicles on the street at the same time. As a result, with a low-cost camera set up on the street, the rate of PM2.5 can be estimated directly without any help from specific particle sensors and promptly warn the people. The research is divided into specific sections as follow:

Design a software system which acquires video tracking traffic density and preprocesses corresponding PM2.5 signal. A survey of state-of-the-art object detection algorithms for vehicle tracking implementation in Vietnam was conducted.

Develop a vehicle tracking algorithm based on deep-learning image classification backbones and background subtraction for detection proposal.

Qualitatively show the correlation relationship between PM2.5 density measured by low-cost sensor and traffic density detected by the traffic tracking algorithm.

Develop a mathematical model to estimate the PM2.5 density based on the traffic density detected from tracking algorithm. The estimation result showed a correlation with observed PM2.5 signal ground-truth acquired from PM2.5 measurement device.

## II. RELATED WORKS

### A. Sensor PMS7003

To respond to the emerging air pollution problem, a lot of efforts have been made to design a low-cost air quality measuring device for better scalability. Therefore, in our study, we used PMS7003 sensor to build our air pollution measuring device and used it as ground-truth for its affordability and outstanding performance compared to its other competitor in the aspects of performance, durability, and its high linear fitting coefficient [4]. The sensor measures PM1.0, PM2.5, and PM10 based on the light scattering principle, with which light scattered by the particles in the atmosphere is detected and its density is measured to interpret the corresponding particle density. In our work, PM2.5 data was collected from the PMS7003 sensor and used as ground truth to evaluate our PM2.5 estimation inferenced from the traffic density detected and tracked with our vehicle tracking system implemented with object detection algorithms. Therefore, we also conducted a search for a suitable objection detector for our vehicle tracking task.

### B. Object detection models

The Mask-RCNN is a deep-learning model with competitive performance on object detection and segmentation tasks. The model was developed based on the two-stage detection model which includes a region proposal stage and classification stage based on the features extracted from the proposed region. The model showed a better performance than the winner in COCO 2016 challenge [2]. This suggests its potential for implementation in vehicle tracking task. Nevertheless, Mask-RCNN has an apparent drawback in processing time due to its complex architecture. To overcome the processing time drawback, the single-shot detector, or SSD, was proposed in [3] by using the one-stage detection architecture. The SSD models showed a comparable performance compared to contemporary two-stage models such as Mask-RCNN and Faster-RCNN; however, it also provided a faster process than two-stage rivals. YOLO (You Only Look Once) was also a breakthrough in the object detection field as it was also the first single-stage real-time object recognition algorithm approach. The model was first described in [1]. We used Yolov5 as a candidate for the vehicle recognition task on our video dataset captured on the street. The model performed well under natural conditions, but it becomes a challenge for it when it turns to a heavy-environment case. The Yolov5 model struggled to perform the detection on the low-light environment combined with low-resolution from low-cost video. Moreover, in some cases, the vehicles are moving fast, which causes the blur effect and brings more problems for the model performance.

### C. Vehicle to PM2.5 algorithm

A study from Bangkok Thailand in 2018 [9] proposed the same idea with us in terms of analyzing the relationship between PM2.5 level as related to traffic flow on multiple roadsides using both linear regression and path analysis. The result from statistical model research indicates the direct effect from the flow of traffic on particulate levels in both open and covered areas. The study also showcased that the result from path analysis is more accurate and efficient than linear regression at the low or high PM concentration level.

A guidebook from the European Environment Agency [10] also discussed the exhaust emissions such as PM2.5, N2O, CO2, NH3, etc. from the road transportation based on many vehicle-related factors. The total distance driven by the vehicles, the technology-specific machine to measure the emission factors, and the number of vehicles driven on the street are both formulated together to estimate the level of associated emission more precisely. The study also outputted many result tables including the estimated emission factor values inferred from each specific commercial vehicle technology in Europe.

### D. Emission factors

Apart from PM2.5, there are plenty of emission factors that can be forecasted together to conclude trusted announcements about the current meteorological conditions in the city. A study from California Air Resource Board in 2020 [11] provided the relationship between different vehicle model technologies and their pollutant emission such as NOx, PM2.5, CO, ROG. The emission factors are calculated using US EPA's Compilation of Air Pollutant Emission Factors, Vol. 5 (AP-42, Chapter 13.2.1, Jan. 2011), and ARB's Miscellaneous Process Methodology 7.9, Entrained Paved Road Travel, Paved Road Dust (updated Jan. 2013). The speed factors from vehicles are also considered to estimate the pollutant as well.

## III. METHODS

### A. Processing PM2.5 data

After plotting the raw PM2.5 data, outliers can be observed (fig 3.1). They affect the visual assessment heavily, while removing them would not cause any drastic changes to the end result. Therefore, they are removed using the interquartile range filter. Then, the missing data are linearly interpolated, and the curve is smoothed out via resampling with 1 hour frequency and STL method [13].

*B. Vehicle counting algorithm*

In order to find correlation between PM2.5 concentration and traffic density, counting vehicles is necessary. In order to satisfy this urge, a vehicle counting algorithm is deployed, which utilizes a modified version of background subtraction and a classification model called DenseNet-121 [12].

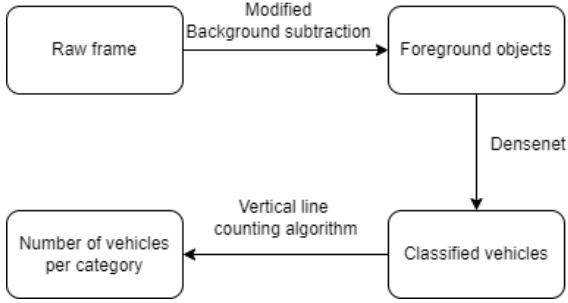

Fig. 1. Flow of vehicle counting algorithm.

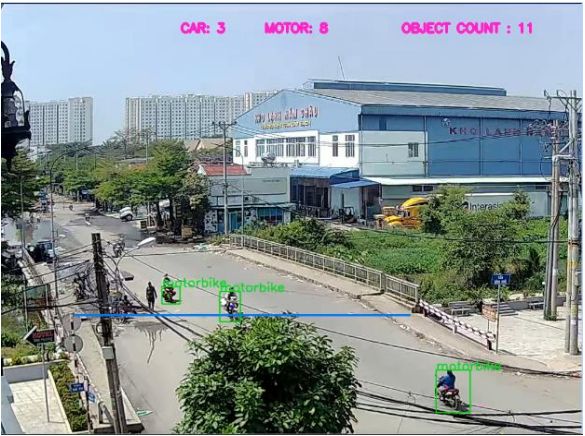

Fig. 2. Demonstration of our tracking system in low-cost camera.

The core of the background subtraction is the algorithm Background Subtractor CNT [14], which, instead of doing just background subtracting to get the background, counts the number of frames that a pixel does not change. If this number exceeds a threshold, the pixel belongs to the background. Before contouring, the resulting frame from background subtraction goes through a post process, in which a series of morphological transformations are applied. To be more specific, the resulting frame is eroded then dilated by a kernel with the size 5 by 5 filled with 1s. Then the kernel of size 5 by 5 in an ellipse shape is applied in MORPH_CLOSE manner (dilation, and then erosion). It is observed that by doing this post process in this order, the contouring works well on the output image. After contouring, the bounding boxes and centers of each moving object are extracted. The bounding box helps to crop the object and feed it to the classification model, while the center helps to count the vehicle, since a vehicle is represented by its center.

For classification, DenseNet-121 is chosen for having a decent accuracy and still suitable for a real time object detection system. The model is transfer-learned on our own dataset, which has 2 classes, namely car and motorbike, and has around 1000 images for training and testing in total.

To count the number of vehicles, a virtual line method is used. In this method, a static line is virtually placed on the frame; for every center that passes this line, the number of the category in which the vehicle belongs increases. With the naïve implementation, a vehicle is likely to be missed, since the sampling rate of the camera is low, about 24 fps. Therefore, the hitbox of the line must be bigger in order to avoid this. However, since the line is bigger, some vehicles might be counted more than one times. Thus, an additional algorithm is developed. For every center that is registered by the line, a rectangle is temporarily drawn. If a center enters the rectangle, it will not be counted. The rectangle will disappear after a certain number of frames not detecting any center. The tracking algorithm pipeline and its in-field demonstration are shown in Fig.1 and Fig.2 respectively.

*C. PM2.5 converting formula*

From Fig. 8, it can be seen that there is a correlation between the PM2.5 concentration and traffic density. However, it is only qualitative. To actually prove they are correlated on paper, Granger causality test is used. Null hypothesis here is defined as follows: traffic density (factor X) does not help to predict future value of PM2.5 concentration (factor Y), i.e., traffic density does not Granger-cause PM2.5 concentration. The result of the granger-causality test is discussed in section 4.

There exists a formula to calculate the amount of PM2.5 emitted by vehicle, and it is stated as follow [15]:

$$E_m = N_m \times EF_m \times VKT_m, \quad (1)$$

where:

$E_m$: mass of emitted PM2.5 (g)

$N_m$: number of vehicles type *m*

$EF_m$: emission factor of vehicle type *m* (g*km$^{-1}$)

$VKT_m$: length of the recorded street segment (km)

The emission factor can be found in California Air Resources Board Emission Factor Table [11]. From formula (1), PM2.5 concentration can be calculated by dividing Em by a volume. Here, it is assumed that PM2.5 particles are uniformly distributed in a cube with one side of length equal to VKTm. Therefore:

$$C_m = E_m \times 10^6 / (VKT_m \times 1000)^3,$$

where:

$C_m$: PM2.5 concentration from vehicle (μg/m$^3$)

*D. Total PM2.5 estimation algorithms*

Intuitively, we know that PM2.5 emitted from vehicles cannot be picked up by the sensor immediately, but rather after some lag. In order to find this lag, we derive a method that will calculate the

mean lag of all local lags. The pseudocode for the method is as follow:

| | Algorithm: Find lag between two time series |
|---|---|
| 1 | X: derived PM2.5 concentration (datetime as |
| 2 | index) |
| 3 | Y: PM2.5 read from sensor (datetime as |
| 4 | index) |
| 5 | Xmins = find local min (X) |
| 6 | Ymins = find local min (Y) |
| 7 | Xmaxs = find local max (X) |
| 8 | Ymaxs = find local max (Y) |
| 9 | j = 0 |
| 10 | Local_lags = [] |
| 11 | FOR i in range(len($X_{mins}$)): |
| 12 |   If j >= len($Y_{mins}$): |
| 13 |     break |
| 14 |   If $X_{mins}$.index[i]<$Y_{mins}$.index[j]: |
| 15 |     Local_lags.append($Y_{mins}$.index[j] - |
| 16 | $X_{mins}$.index[i]) |
| 17 |   j+=1 |
| 18 | j = 0 |
| 19 | FOR i in range(len($X_{maxs}$)): |
| 20 |   If j >= len($Y_{maxs}$): |
| 21 |     break |
| 22 |   If $X_{maxs}$.index[i]<$Y_{maxs}$.index[j]: |
| |     Local_lags.append($Y_{maxs}$.index[j]- $X_{maxs}$.index[i]) |
| |   j+=1 |
| | Lag = mean (Local_lags) |

Vehicle is a source of PM2.5, but it is not the only source. Therefore, in order to derive the total PM2.5 concentration, we need to guess the PM2.5 contribution from other sources. Here, we assume that for the same hour, PM2.5 contributions from other sources are the same, which means we will take one day, calculate the pattern of that day, and use the captured pattern for the following days. To get the pattern, we do the following:

- From the same X and Y as above, we get $X_{shifted}$ by shifting X to the right by the lag.
- We calculate P = Y - $X_{shifted}$, where P is the targeted pattern.

To derive the total PM2.5 concentration of the next day, we add the pattern with PM2.5 concentration from vehicles.

## IV. RESULTS

### A. PM2.5 data processing

The following figure shows how the data look like after each step (from top to bottom):

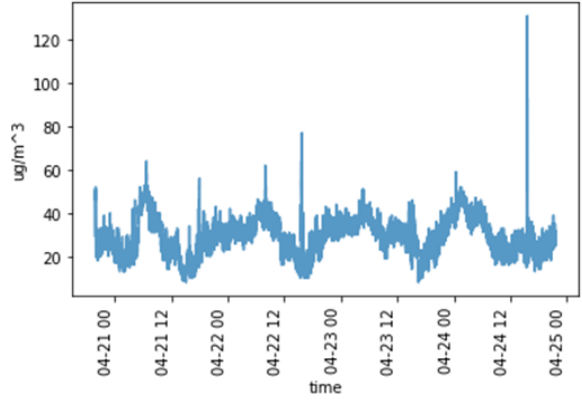

Fig. 3. Raw PM2.5 data.

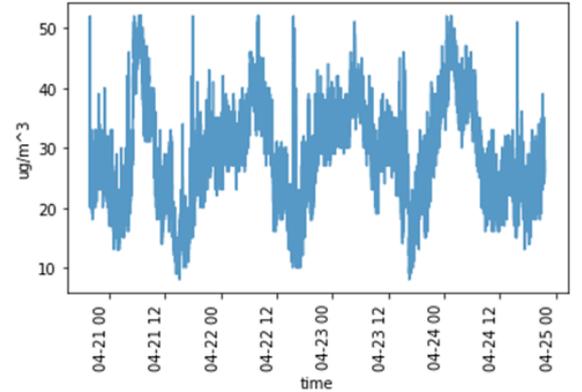

Fig. 4. Filtered PM2.5 data.

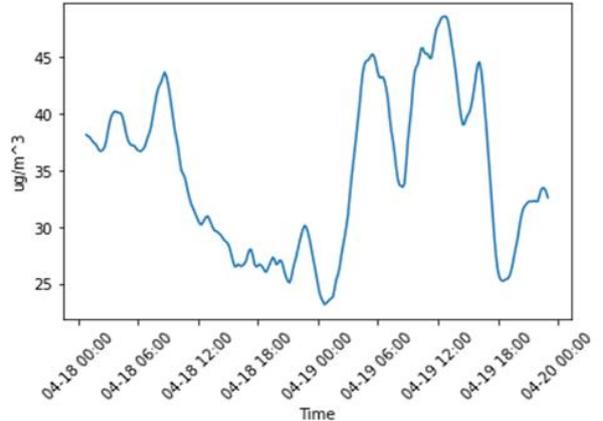

Fig. 5. Interpolated, resampled and smoothened data.

### B. Benchmark

We benchmark our 2-stage object detection model by comparing it with Mask-RCNN, SSD, Yolo Family on AICity challenge 2022 dataset [16].

| Model Names | AUC | FPS |
|---|---|---|
| **Modified background Subtraction** | 0.3089 | 28.24 |
| Mask RCNN | 0.1477 | 3.94 |
| **Yolov3** | **0.4** | **28.89** |
| Yolov5s | 0.36 | 28.103 |
| Yolov5m | 0.37 | 21.659 |
| Yolov5l | 0.34 | 20.107 |
| **Yolov7-tiny** | **0.37** | **49.090** |

| Yolov7 | 0.42 | 33.297 |
|---|---|---|
| SSD512 | 0.45 | 13.7 |

Fig. 6. Comparison of AUC and FPS between different model and algorithm

Here, we compare using 2 metrics, namely area under precision-recall curve (AUC) and processed frames per second. As we can see, the yolo families out-perform other models, except SSD512 since it has the highest AUC. However, through actual usage, yolo does not work well at night, as can be seen in the following graph:

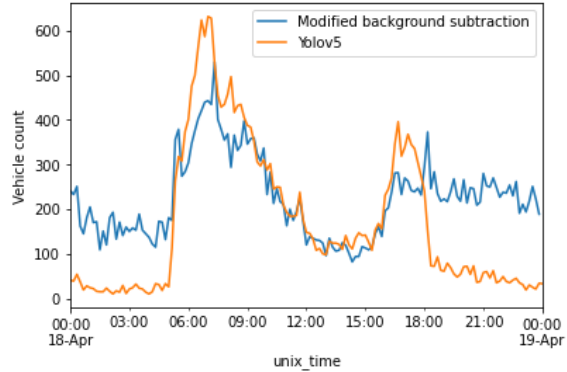

Fig. 7. Plot number of vehicle using Yolo and Modified background subtraction.

### C. Correlation between traffic density and PM2.5 (qualitative)

After preprocessed the vehicle counting data and PM2.5 data read from sensor, we have the following plot:

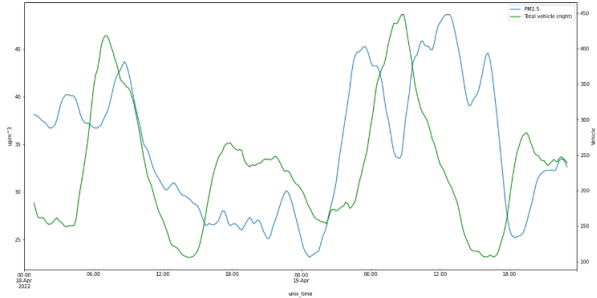

Fig. 8. Plot number of vehicle and PM2.5.

It can be seen from the plot that vehicle density does have some correlation with PM2.5. The lag between the two signals can also be observed.

### D. PM2.5 estimation result

|  | Car count | Motorbike count | Total vehicle count |
|---|---|---|---|
| PM2.5 from sensor | 0.0002 | 0.0007 | 0.0052 |

Fig. 9. p-values of Granger causality test.

The p-values in all three cases are lower than 5%, null hypothesis is rejected, which means traffic density Granger-cause PM2.5 concentration.

After getting the lag, we calculate the PM2.5 concentration of the next day using only information about the vehicle and plot it along with the PM2.5 concentration from the sensor:

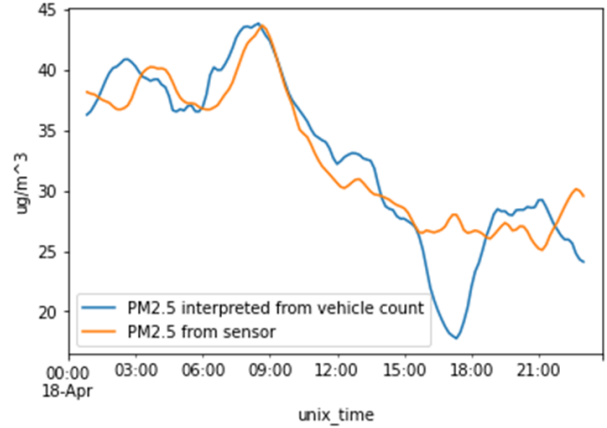

Fig. 10. Line plot derived PM2.5 and PM2.5 read from sensor.

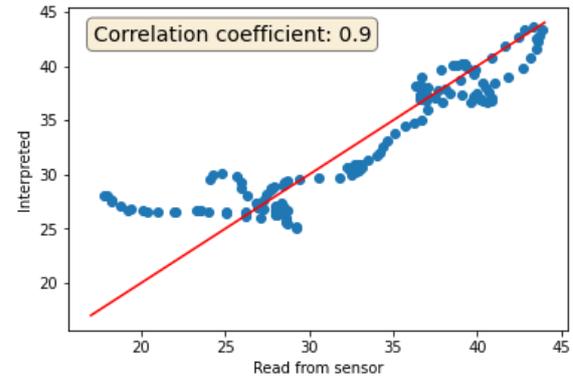

Fig. 11. Scatter plot (correlation plot) of derived PM2.5 and PM2.5 read from sensor.

The two plots show that the derived PM2.5 concentration is highly correlated with PM2.5 read from sensor (ground truth). Here, Pearson correlation coefficient is 0.9.

## V. CONCLUSIONS

In this work, we proposed a low-lost PM2.5 estimation method based on a traffic surveillance system in urban areas. We first investigated and developed our vehicle detection and tracking algorithms to satisfy both performance and processing time criteria. Our proposed detection method shows a comparable performance compared to other state-of-the-art methods based on AUC and fps evaluation metrics. YOLO family models showed the most outstanding performance in benchmark dataset, but it was shown to have some difficulty to train and implement YOLO models with the expected efficiency for the problem in Vietnam, while our method had a robust performance for the in-field implementation and showed an advantage in scalability. The observed traffic density from the tracking system demonstrated qualitatively a closed relation with the measured PM2.5 data. This relationship was further investigated with a mathematical model, which can be used to estimate PM2.5 values from the traffic density data acquired from the tracking system. Therefore, the relation between the two data was quantitatively analyzed, and the estimation result had a great correlation with the measured PM2.5 data. Moreover, by using Granger causality, we pointed out the correlation between the estimation result and the measured PM2.5 ground-truth.